%% file: main.tex
\documentclass[reprint,prl,twocolumn,superscriptaddress,showpacs,showkeys,floatfix,preprintnumbers]{revtex4-2}

\pdfoutput=1

\usepackage{braket}
\usepackage{xargs}
\usepackage{extarrows}
\usepackage{mathtools}
\usepackage[english]{babel}
\usepackage[utf8]{inputenc}
\usepackage{amsmath,amssymb,braket,slashed,mathrsfs}
\usepackage{pifont}
\usepackage{MnSymbol}
\usepackage{graphicx}
\usepackage{color,hyperref}
\usepackage{multirow,array}
\usepackage{ulem}

\usepackage{amsmath} 
\usepackage{graphicx}
\usepackage{booktabs}
\usepackage{multirow}
\usepackage{subfigure}
\usepackage{float}
\usepackage{bm}
\usepackage{diagbox}
\usepackage{multirow}
\usepackage{tabularx}

\usepackage{cleveref}

\usepackage[utf8]{inputenc}
\usepackage[english]{babel}

\input{values2.tex}

\usepackage{graphicx} 
\usepackage{float}
\graphicspath{ {./images/} }
\usepackage{braket}
\usepackage{dcolumn}

\newcommand{\xfs}[2]{{#2}}

\begin{document}

\title{Lattice QCD calculation of the pion mass splitting}
\author{Xu Feng}
\email{xu.feng@pku.edu.cn}
\affiliation{School of Physics and State Key Laboratory of Nuclear Physics and Technology, Peking University, Beijing 100871, China}
\affiliation{Collaborative Innovation Center of Quantum Matter, Beijing 100871, China}
\affiliation{Center for High Energy Physics, Peking University, Beijing 100871, China}
\author{Luchang Jin}
\email{ljin.luchang@gmail.com}
\affiliation{Department of Physics, University of Connecticut, Storrs, CT 06269, USA}
\affiliation{RIKEN-BNL Research Center, Brookhaven National Laboratory, Building 510, Upton, NY 11973}
\author{Michael Joseph Riberdy}
\email{michaeljosephriberdy@gmail.com}
\affiliation{Department of Physics, University of Connecticut, Storrs, CT 06269, USA}
\date{\today}

\begin{abstract}
    We use the infinite volume reconstruction method to calculate the charged/neutral pion mass difference. The hadronic tensor is calculated using lattice QCD and then combined with an analytic photon propagator, and the mass splitting is calculated with exponentially-suppressed finite volume errors.
The calculation is performed using
six gauge ensembles generated with $2+1$-flavor domain wall fermions,
and five ensembles are at the physical pion mass.
Both Feynman and Coulomb gauges are adopted in the calculation and agree well when the lattice spacing approaches zero.
After performing the continuum
extrapolation and examining the residual finite-volume effects, 
we obtain the pion mass splitting
$\Delta m_\pi = 4.534(42)_\text{stat}(43)_\text{sys}~\mathrm{MeV}$,
which agrees well with experimental measurements.

\end{abstract}

\maketitle

\section{Introduction}

\xfs{}{One of the central goals in high energy physics is understanding the nature of the matter that we observe in the universe.
As one of the four known fundamental interactions, the strong interaction binds together quarks and gluons into hadrons and most of the hadron mass 
arises in turn from the binding energy;
the individual quarks provide only a very small portion of the mass.
Precision calculation based on the lattice formulation of quantum chromodynamics (QCD), the theory of the strong interaction, can manifest
its success by computing the various hadron spectra, which agree well with experimental measurements~\cite{Kronfeld:2012uk}.
When the precision reaches percent or subpercent level, another fundamental force, electromagnetic interaction, 
is urged to be considered in theoretical calculations, although its effects are suppressed by a 
factor of the fine-structure constant $\alpha_{\mathrm{EM}}\approx 1/137$. Inclusion of the electromagnetic corrections does not only 
provide precise hadron spectroscopy~\cite{Blum:2007cy,Blum:2010ym,Ishikawa:2012ix,deDivitiis:2013xla,Borsanyi:2014jba,Horsley:2015eaa,Horsley:2015vla,Giusti:2017dmp,Boyle:2017gzv,MILC:2018ddw,CSSM:2019jmq},
but also plays an important role in studies of leptonic and semileptonic decays of hadrons~\cite{Carrasco:2015xwa,Lubicz:2016xro,Giusti:2017dwk,DiCarlo:2019thl,Feng:2020zdc,Christ:2020jlp,Ma:2021azh}, 
which dramatically expands the horizon of lattice QCD studies.}

\xfs{}{Among various hadrons, 
pions play a unique role in the development of theoretical particle and nuclear physics. Pions were first proposed by H. Yukawa in 1935 as the carrier particles 
of the strong nuclear force~\cite{Yukawa:1935xg}. As Nambu-Goldstone bosons~\cite{Nambu:1960tm,Goldstone:1961eq}, pions result from the spontaneous break down of chiral symmetries of QCD effected by quark condensation 
and serve as the active degrees of freedom sensitive to chiral dynamics~\cite{Weinberg:1978kz,Gasser:1982ap,Gasser:1983yg}. Besides, the anomalous decay rate of the neutral pion led to the 
discovery of Adler-Bell-Jackiw anomaly of quantum electrodynamics (QED)~\cite{Adler:1969gk,Bell:1969ts}, which revealed for the first time the violation of classical symmetry by quantum corrections.
It can be concluded that the thorough study of the nature of the pions is a key to our better understanding of QCD and the 
strong interaction.
In this work, we focus on the study of the mass splitting between the charged and neutral pions, which represents the interplay between 
two fundamental interactions, the strong and the electromagnetic. For two reasons, the pion mass splitting is ideal for a lattice QCD calculation and 
for an exploratory study of new methodology. First, pions are the lightest hadrons and their correlation functions have very good statistical signals. 
Second, the isospin breaking effects of the up and down quark mass difference are suppressed by a 
factor of $(m_u-m_d)^2/\Lambda_{\mathrm{QCD}}^2\sim 10^{-4}$ with $m_{u/d}$ the up/down quark mass and $\Lambda_{\mathrm{QCD}}\approx 350$ MeV the nonperturbative QCD scale,  
leaving the electromagnetic effects the leading contribution to the pion mass splitting. Thus the ambuguity in separating the isospin breaking effects 
from the electromagnetic interaction and the quark mass difference
becomes irrelevant in this study. One can simply perform the lattice QCD calculation in the isospin symmetric limit and compute the four-point correlation functions for the QED self-energy 
diagrams.}

\xfs{}{
In practice we adopt the infinite-volume reconstruction (IVR) method proposed in Ref.~\cite{Feng:2018qpx}, which allows us to calculate electromagnetic corrections to stable hadron masses
with only exponentially suppressed finite-volume effects. Upon its development, this method has been successfully applied and extended to various electroweak processes involving photon or massless leptonic propagators~\cite{Tuo:2019bue,Feng:2019geu,Christ:2020jlp,Christ:2020hwe,Tuo:2021ewr}.
This is the first time
that we apply this methodology to the calculation of the pion mass splitting. 
The calculation includes the complete diagrams with both connected and disconnected quark-field contractions.
We employ two gauge fixings for the photon propagator and confirm that the lattice results are consistent in the continuum limit.
We utilize the random field sparsening technique proposed in Ref.~\cite{Li:2020hbj}, which allows us to improve the precision of the correlation functions significantly with only
a modest cost of computational resources. By using five gauge ensembles generated with $N_f=2+1$ domain wall fermions at physical pion mass and one additional ensemble at 
$m_\pi\approx 340$ MeV, we obtain the pion mass splitting with a percent-level uncertainty, which is about 5-10 times smaller than previous
lattice QCD calculations~\cite{Gupta:1984tb,Gupta:1987zc,deDivitiis:2013xla,Horsley:2015eaa,Giusti:2017dmp}. (See Table~\ref{tab:previous_results}.
Refs.~\cite{Gupta:1984tb,Gupta:1987zc} present the pioneering quenched calculations and thus the results are not included in Table~\ref{tab:previous_results}.)
For the first time in the literature,
we have clearly resolved and included the contribution from the quark disconnected diagram
to the pion mass splitting
(see Eq.~(\ref{eq:discon-corr}) and the diagram below this equation).
This diagram is related to the $\pi^0-\eta-\eta'$ mixing and has also been calculated in Ref.~\cite{CSSMQCDSFUKQCD:2021rvs}.
}

\begin{table}[]
\begin{tabular}{|c|c|}
\hline
Reference & $m_{\pi^{\pm}}-m_{\pi^{0}} (\text{MeV})$ \\ \hline
RM123~2013~\cite{deDivitiis:2013xla} & $5.33(48)_\text{stat}(59)_\text{sys}$
\footnote{Converted from $m_{\pi^\pm}^2 - m_{\pi^0}^2 = 1.44(13)(16)\times10^3~\text{MeV}^2$.} \\ \hline
R.~Horsley et al.~2016~\cite{Horsley:2015eaa} & $4.60(20)_\text{stat}$ \\ \hline
RM123~2017~\cite{Giusti:2017dmp} & $4.21(23)_\text{stat}(13)_\text{sys}$ \\ \hline
This work & $4.534(42)_\text{stat}(43)_\text{sys}$ \\ \hline
\end{tabular}
\caption{
Previous lattice calculations of $m_{\pi^{\pm}}-m_{\pi^{0}}$ are compared to this work.
}
\label{tab:previous_results}
\end{table}

\section{ Infinite Volume Reconstruction Method}
The hadron mass extraction
relies on the calculation of the hadron QED self-energy for a stable hadronic state $N$ via the following infinite volume Euclidean space-time integral
\begin{equation}
\Delta M =
\mathcal{I}=
\frac{1}{2}\int d^4x\,\mathcal{H}_{\mu,\nu}(x) S^\gamma_{\mu,\nu}(x),
\label{eq:init}
\end{equation}
where the hadronic part $\mathcal{H}_{\mu,\nu}(x)=\mathcal{H}_{\mu,\nu}(t, \vec x)$ is given by
\begin{equation}
\mathcal{H}_{\mu,\nu}(x)
=\frac{1}{2 M}\langle N(\vec 0)|T [J_\mu(x) J_\nu(0)]|N(\vec 0)\rangle,
\label{eq:def-H}
\end{equation}
where $|N(\vec p)\rangle$ indicates a hadronic state $N$ with mass $M$ and spatial momentum $\vec p$,
$J_\mu=2e\, \bar{u}\gamma_\mu u/3 - e\, \bar{d}\gamma_\mu d/3 - e\, \bar{s}\gamma_\mu s/3$
is the electromagnetic current,
and $S^\gamma_{\mu,\nu}$ is the photon propagator whose form is analytically known.
In Ref.~\cite{Feng:2018qpx}, we introduced the IVR method to relate
the infinite volume integration of infinite volume hadronic matrix elements
$\mathcal{H}_{\mu,\nu}$ to
a finite lattice volume integration of finite volume matrix elements
$\mathcal{H}^L_{\mu,\nu}$
with only exponentially suppressed finite volume errors.
This is accomplished with the following three steps:
\begin{enumerate}
\item  We pick $t_s$ to separate the infinite volume integral into
two parts:
\begin{equation}
\label{eq:separate_short_long}
\mathcal{I} = \mathcal{I}^{(s)}+\mathcal{I}^{(l)} \\
\end{equation}
where $\mathcal{I}^{(s)}$ and $\mathcal{I}^{(l)}$
are the ``short distance'' ($|x_t| < t_s$) and ``long distance'' ($|x_t| \ge t_s$) contributions respectively.
\item For sufficiently large $t_s$, $\mathcal{I}^{(l)}$ is dominated by
the lightest single particle intermediate states, and can be calculated using
the hadronic matrix elements at fixed time separation $t_s$.
The excited-state effects ignored in this step are exponentially suppressed by large $t_s$.
\item The next step is to approximate $\mathcal{H}_{\mu,\nu}$ using $\mathcal{H}^L_{\mu,\nu}$
and restrict the integration region to the finite lattice volume.
This step only introduces exponentially suppressed finite volume errors since
we only need to use $\mathcal{H}^L_{\mu,\nu}(x)$ for $|x_t| \le t_s \lesssim L$.
\end{enumerate}
The final formula obtained in Ref.~\cite{Feng:2018qpx} is expressed in terms of lattice-calculable quantaties:
\begin{eqnarray}
\mathcal{I}^{(s)}\approx\mathcal{I}^{(s,L)}&=&\frac{1}{2}\int_{-t_s}^{t_s}\!\! dt\int_{-L/2}^{L/2}\!\! d^3\vec{x}\, \mathcal{H}^L_{\mu,\nu}(\vec{x}) S^{\gamma}_{\mu,\nu}(\vec{x}) \\ 
\mathcal{I}^{(l)}\approx\mathcal{I}^{(l,L)}&=&\int_{-L/2}^{L/2}\!\! d^3\vec{x}\, \mathcal{H}^L_{\mu,\nu}(t_s,\vec{x}) L_{\mu,\nu}(t_s,\vec{x})
\end{eqnarray}
where $L_{\mu\nu}(t_s,\vec{x})$ is an infinite volume QED weighting function, defined as:
\begin{eqnarray}
&&L_{\mu,\nu}(t_s,\vec{x}) =
\int\!\!\frac{d^3p}{(2\pi)^3}
e^{i \vec{p}\cdot \vec{x}}
 \int_{t_s}^\infty\!\! dt\,
e^{-(E_{\vec{p}}-M)(t-t_s)}
\nonumber
\\ && \hspace{3cm} \times
 \int\! d^3\vec{x}'\,
e^{-i \vec{p}\cdot \vec{x}'}
S^\gamma_{\mu,\nu} (t,\vec{x}').
\end{eqnarray}
In this paper, we use $t_{s}=L / 2$ for our final result to ensure both the excited states and finite volume errors introduced in steps 2 and 3 described above are exponentially suppressed as we increase
the lattice size $L$.

\subsection{Gauge-Specific Expressions}
We also present the relevant expressions for the photon propagator $S^{\gamma}_{\mu,\nu}(x)$ and QED weighting function $L_{\mu,\nu}(t_s,\vec{x})$ in Feynman and Coulomb Gauges:
\begin{eqnarray}
S_{\mu,\nu}^{\gamma,\text{Feyn}}(x)
=\delta_{\mu,\nu} \int\!\!\frac{d^4p}{(2\pi)^4}\,\frac{e^{ipx}}{p^2}
= \frac{\delta_{\mu,\nu}}{4\pi^2x^2}
\end{eqnarray}
\begin{eqnarray}
L_{\mu,\nu}^{\text{Feyn}}(t_s,\vec{x})
&=&
\frac{\delta_{\mu,\nu}}{2\pi^2}
\int_0^\infty\!\!dp\,
\frac{\sin(p|\vec{x}|)}{2(p+E_p-M)|\vec{x}|} 
e^{-p t_s}
\end{eqnarray}
\begin{eqnarray}
        \label{eq:coulomb_photon_propagator}
         &&
         \hspace{-0.3cm}
         S_{\mu,\nu}^{\gamma,\text{Coul}}(x_t,\vec x)
         = \frac{\delta_{\mu,t}\delta_{\nu,t}}{4\pi|\vec{x}|}\delta(x_t) \\
         && + \nonumber \frac{(1-\delta_{\mu,t})(1-\delta_{\nu,t})}{2(2\pi)^{3}}\int(\delta_{\mu,\nu}-\frac{p_{\mu}p_{\nu}}{\vec{p}^{2}})\frac{e^{-|\vec{p}|x_t+i\vec{p}\cdot\vec{x}}}{|\vec{p}|}d^{3}p
         \\
         &=&
         \frac{\delta_{\mu,t}\delta_{\nu,t}}{4\pi|\vec{x}|}\delta(x_t)
         + \frac{(1-\delta_{\mu,t})(1-\delta_{\nu,t})}{(2\pi)^{2}}
         \\ \nonumber
         &&
         \times\int_{0}^{\infty}
         \frac{e^{-p x_t}}{|\vec{x}|}
         \Bigg[
         \delta_{\mu,\nu}
         \Big{(}
         \sin(p|\vec{x}|) +
         \frac{\cos(p|\vec{x}|)}{p|\vec{x}|}-\frac{\sin(p|\vec{x}|)}{p^2|\vec{x}|^{2}}
         \Big{)} \\
         \nonumber
         && \hspace{1cm} + \frac{x_\mu x_\nu}{|\vec x|^2}
         \Big{(}
         -\sin(p|\vec{x}|)-\frac{3\cos(p|\vec{x}|)}{p|\vec{x}|}+\frac{3\sin(p|\vec{x}|)}{p^2|\vec{x}|^{2}}
         \Big{)}
         \Bigg]
         dp
\end{eqnarray}
\begin{eqnarray}
L_{\mu,\nu}^{\text{Coul}}(t_s,\vec{x})
&=&
0
\end{eqnarray}
At the origin ($x=0$), the continuum photon propagator is divergent.
In our lattice calculation, we regularize this divergence with the following choice:
\begin{eqnarray}
\label{eq:photon_prop_origin_feyn}
S_{\mu,\nu}^{\gamma,\text{Feyn}}(0) &=&
\frac{\delta_{\mu,\nu}}{a^8}
\int_{-a}^a d x
\int_{-a}^a d y
\int_{-a}^a d z
\int_{-a}^a d t
\\&&
\nonumber
\hspace{1cm}
\times
\frac{(a-|x|)(a-|y|)(a-|z|)(a-|t|)}{4\pi^2(x^2 + y^2 + z^2 + t^2)}
\\
&=& 2.76963~\delta_{\mu,\nu}/ (4\pi^2 a^2)
\end{eqnarray}
\begin{eqnarray}
\label{eq:photon_prop_origin_coul}
S_{t,t}^{\gamma,\text{Coul}}(0) &=&
\frac{1}{a^7}
\int_{-a}^a d x
\int_{-a}^a d y
\int_{-a}^a d z
\\&&
\nonumber
\hspace{1cm}
\times
\frac{(a-|x|)(a-|y|)(a-|z|)}{4\pi\sqrt{x^2 + y^2 + z^2}}
\\
&=& 1.88231 / (4\pi a^2)
%
\end{eqnarray}

\section{Relevant Diagrams}
Contributions to the charged/neutral QED pion mass splitting are derived from the following hadronic matrix element: 
\begin{equation}
\mathcal{H}_{\mu,\nu}^{\pm,0}(x)
=\frac{1}{2 M_{\pi}}\langle \pi^{\pm,0}(\vec 0)|T [J_\mu(x) J_\nu(0)]|\pi^{\pm,0}(\vec 0)\rangle 
\end{equation}
where $\pi(\vec{0})$ represents a pion with zero momentum.
Only two contractions contribute to this matrix element~\cite{deDivitiis:2013xla}.
One yields (where $S$ represents the light quark propagator)
\begin{eqnarray}
\label{eq:discon-corr}
&&C^{1}_{\mu,\nu}(x-y) =
\big\langle\text{Tr} \big(S (x; t_{\text{src}}) \gamma_5 S (t_{\text{src}} ; x) \gamma_{{\mu}}\big)
\\
\nonumber
&&
\hspace{2.5cm}\times
\text{Tr} \big(S (y ; t_{\text{snk}}) \gamma_5 S (t_{\text{snk}} ; y) \gamma_{\nu}\big)\big\rangle_{\text{QCD}}
\end{eqnarray}
which is related to the following quark disconnected diagram.
\begin{center}
    \includegraphics[scale=1.5]{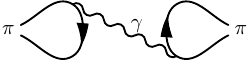}
\end{center}
The other possible contraction yields (up to the insertion of a photon propagator)
\begin{eqnarray}
\label{eq:con-corr}
&&C^{2}_{\mu,\nu}(x-y) =
\big\langle\text{Tr} \big(\gamma_{{\mu}} S (x ; t_{\text{src}}) \gamma_5 S (t_{\text{src}} ; y)
\\
\nonumber
&&\hspace{2.5cm}
\times
\gamma_{\nu} S (y ; t_{\text{snk}}) \gamma_5 S (t_{\text{snk}} ; x)\big)\big\rangle_{\text{QCD}}
\end{eqnarray}
which is represented by the following quark connected diagram.
\begin{center}
    \includegraphics[scale=1.5]{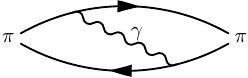}
\end{center}
To ensure projection onto the pion state, we fix the time separation between the pion interpolating operators
and the closest electromagnetic current operators to be
$t_\text{sep} = t_\text{snk} - x_t = y_t - t_\text{src}$
for both diagrams
(assuming $x_t \ge y_t$).
The values of $t_\text{sep}$ for each ensemble are listed in Table~\ref{tab:latinfo}.
Coulomb gauge fixed wall source is used for the pion interpolating operators.
Combining these diagrams yields the hadronic contribution to the mass shift, which is represented by 
\begin{eqnarray}
    &&
    \hspace{-0.8cm}
    \mathcal{H}_{\mu,\nu}^{\pm}(x) - \mathcal{H}_{\mu,\nu}^{0}(x) \\ \nonumber &=& 
    e^{2}
    Z_V^2 L^3
    {
    C^{1}_{\mu,\nu}(x)
    +C^{1}_{\nu,\mu}(-x)
    -C^{2}_{\mu,\nu}(x)
    -C^{2}_{\nu,\mu}(-x)
    \over
    2\, C^\text{AW}_\pi(|x_t| + 2 t_\text{sep})
    },
\end{eqnarray}
where 
\begin{eqnarray}
\label{eq:c_pi}
C^\text{AW}_\pi(t + 2 t_\text{sep})
&=&
C_{\pi}(2t_\text{sep}) e^{-M_\pi t}
\nonumber\\&&\hspace{0.5cm}
\times\big(1-e^{-M_{\pi}(T-4t_{\text{sep}})} \big),
\\
C_\pi(t_2 - t_1)
&=&
\big\langle\text{Tr} \big(
S (t_2;t_1)) \gamma_5 S (t_1;t_2) \gamma_{{5}}
\big)\big\rangle_{\text{QCD}}.
\nonumber
\end{eqnarray}
Local lattice vector current operators are used in the contraction,
and the corresponding renormalization factor $Z_V$ is included in the above formula.
We have taken the around the world effects into account when calculating the
pion correlation function in Eq.~(\ref{eq:c_pi}).
In principle,
both the isospin breaking due to the up / down quark mass difference
and the QED effects described above contribute to the pion mass difference.
For many hadronic observables, the corrections due to
the up / down quark mass difference
are at the same order as the QED corrections.
Therefore, we treat $\mathcal{O}((m_{d}-m_{u}) / \Lambda_\text{QCD})$  corrections to be of the same perturbative order as corrections of order $\mathcal{O}(\alpha_{\text{QED}})$.
For pion mass splitting,
the $\mathcal{O}((m_{d}-m_{u}) / \Lambda_\text{QCD})$ effect is zero.~\cite{deDivitiis:2013xla}
Therefore, to order $\mathcal{O}(\alpha_{\text{QED}}, (m_{d}-m_{u}) / \Lambda_\text{QCD})$, the two diagrams discussed above represent the only contributions to the pion mass splitting.

\section{Numerical Results}
\begin{figure}
    \flushright
    \includegraphics[scale = 1.05]{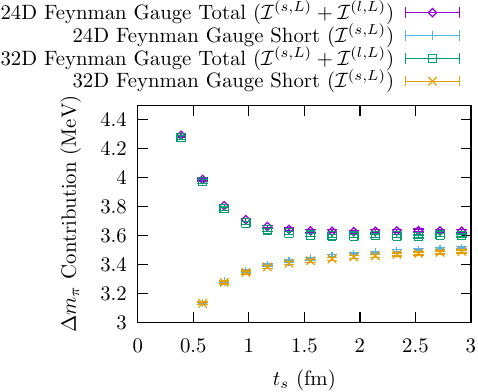}\\
    \includegraphics[scale = 1.05]{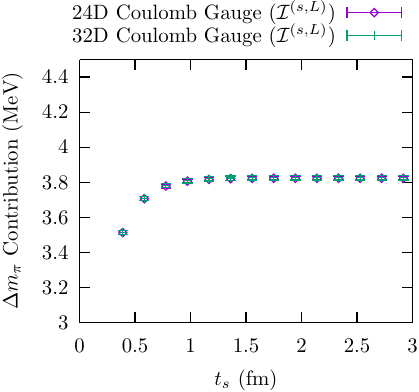}
    \caption{The pion mass shift calculated using the 24D and 32D ensembles is shown as a function of $t_{s}$ for Feynman gauge (top) and Coulomb gauge (bottom).}
    \label{fig:ts_dependence}
\end{figure}

\begin{table}[]
\begin{tabular}{|c|c|c|c|c|c|}
\hline
 & Volume & $a^{-1}$ (GeV) & $L$ (fm) & $M_{\pi}$ (MeV) & $t_\text{sep}$ (a) \\ \hline
48I & $48^{3} \times 96$ & 1.730(4) & 5.5 & 135 & 12 \\ \hline
64I & $64^{3} \times 128$ & 2.359(7) & 5.4 & 135 & 18\\ \hline
24D & $24^3 \times 64$ & 1.0158(40) & 4.7 & 142 & 8\\ \hline
32D & $32^3 \times 64$ & 1.0158(40) & 6.2 & 142 & 8\\ \hline
32Dfine & $32^3 \times 64$ & 1.378(7) & 4.6 & 144 & 10 \\ \hline
24DH & $24^3 \times 64$ & 1.0158(40) & 4.7 & 341 & 8\\ \hline
\end{tabular}
\caption{
List of the ensembles used in the calculations and their properties.
They are generated by the RBC and UKQCD collaborations.~\cite{Blum:2014tka}
Note we use a partially quenched quark mass for 48I and 64I ensembles.
The unitary pion mass for both 48I and 64I ensembles is $139~\mathrm{MeV}$.
We use unitary quark masses for all the other ensembles.
}
\label{tab:latinfo}
\end{table}
\begin{table}[]
\begin{tabular}{|c|l|l|l|}
\hline
 & Feyn (MeV)
     & Coul (MeV)
     & Coul-t (MeV)\\ \hline
48I & $\daf\ $ & $\dac\ $ & $\cgia\ $ \\ \hline
64I & $\dbf\ $ & $\dbc\ $ & $\cgib\ $ \\ \hline
24D & $\dEf\ $ & $\dEc\ $ & $\cgiE\ $ \\ \hline
32D & $\dcf\ $ & $\dcc\ $ & $\cgic\ $ \\ \hline
32Dfine & $\ddf\ $ & $\ddc\ $ & $\cgid\ $ \\ \hline
24DH & $\dff\ $ & $\dfc\ $ & $\cgif\ $ \\ \hline
\end{tabular}
\caption{
Contributions to the pion mass shift are shown by ensemble.
The fourth column displays the Coulomb potential contribution.
The statistical uncertainty is shown in parenthesis.
}
\label{tab:latresults}
\end{table}
Calculation of the hadronic function $\mathcal{H}_{\mu\nu}^{L}(x)$ was performed on six ensembles generated by the RBC and UKQCD collaborations.
The names and attributes of the ensembles are shown in Table \ref{tab:latinfo}.
The properties of these ensembles are studied in detail in Ref.~\cite{Blum:2014tka}.
In particular, the lattice spacing and other basic parameters of the ensemble
are determined by matching the lattice calculation
of the masses of pion, kaon, and the $\Omega$ baryon to their experimental values.
We use All-modes-averaging (AMA) \cite{Blum:2012uh,Shintani:2014vja}, zM\"obius \cite{Mcglynn:2015uwh}, and compressed eigenvector deflation \cite{Clark:2017wom} methods to accelerate
the calculation of the propagators.
Figure \ref{fig:ts_dependence} shows the mass shift $\Delta m_{\pi}\equiv m_{\pi^{+}}-m_{\pi^{0}}$ as a function of $t_{s}$ in the Feynman and Coulomb gauges for the 24D and 32D ensembles.
It can be seen from the plots that the finite volume effects are very small as the differences between 24D and 32D are barely visible.
Also, for $t_s\gtrsim 1.5~\mathrm{fm}$, the final results have a very mild dependence on $t_s$,
suggesting the excited states contribution beyond $t_s$,
which is exponentially suppressed and ignored in the IVR method,
is indeed quite small.
In the following analysis, we stick to:
\begin{eqnarray}
t_s = L / 2.
\end{eqnarray}
With this choice,
the finite volume effects at fixed $t_s$
and the excited intermediate states' contribution beyond $t_s$
are both exponentially suppressed by the spatial lattice size $L$,
and we will refer to the sum of these two effects
as the finite volume effects in the following analysis.
In Feynman gauge, the difference between 32D and 24D is $-0.035(16)~\mathrm{MeV}$.
This is consistent with a scalar QED calculation,
which yields $-0.022~\mathrm{MeV}$~\cite{Feng:2018qpx}.
In Coulomb gauge, the difference between 32D and 24D is $0.002(17)~\mathrm{MeV} $.

The results for each ensemble are presented in Table~\ref{tab:latresults}.
In the table, we also show the Coulomb potential contribution to the pion mass difference.
This contribution comes from the time component of the Coulomb gauge photon propagator, $S^{\gamma,\text{Coul}}_{t,t}$ in Eq.~(\ref{eq:coulomb_photon_propagator}).

\section{Extrapolation to the physical point}
\begin{figure}
    \centering
    \includegraphics[scale=1]{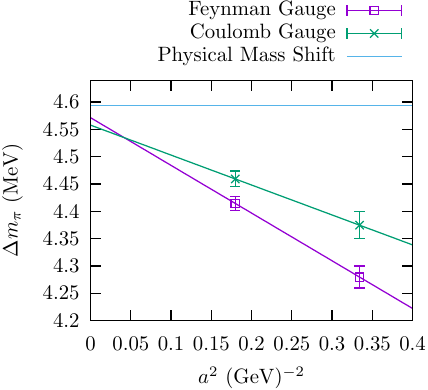}
    \caption{Feynman and Coulomb gauge mass shifts are shown as a function of $a^{2}$ for the Iwasaki ensembles 48I and 64I.
    }
    \label{fig:continuum_limit}
\end{figure}

We perform the continuum extrapolation for the Iwasaki ensembles (48I and 64I)
and the I-DSDR ensembles 24D, 32Dfine, and 24DH separately
using the following formula ($m_{\pi,\text{phys}} = 135~\mathrm{MeV}$).
\begin{eqnarray}
    \label{eq:fit_1}
    m_\pi \Delta m_{\pi}(a^2, m_\pi)&=&m_{\pi,\text{phys}} \Delta m_{\pi}(0, m_{\pi,\text{phys}}) \\ \nonumber
    &&\hspace{-1cm} \times \Big( 1+ c_{1}a^{2} + c_{2}({m_{\pi}}^2-m_{\pi,\text{phys}}^{2})\Big)
\end{eqnarray}
When using Domain wall fermions in lattice calculations,
lattice artifacts, which scale as $\mathcal O(a)$ or $\mathcal O(a^{2n+1})$, are absent~\cite{RBC:2012cbl,Blum:2014tka}.
We choose to include an $\mathcal{O}(a^{2})$ term to account for the lattice artifact.
We also include a naive pion mass dependence term in this extrapolation formula to
accommodate the small pion mass difference in the I-DSDR ensembles.
This pion mass dependence term does not effect our main result,
which is taken from the continuum extrapolation of
the Iwasaki ensembles exactly at the physical pion mass.
The extrapolation results of the Iwasaki 48I and 64I ensembles are shown in Figure~\ref{fig:continuum_limit},
where separate fits were conducted for Feynman and Coulomb gauges.

To estimate the discretization systematic error
for the fitted value of $\Delta m_{\pi}(0,m_{\pi,\text{phys}})$,
we perform two slightly different fits by:
\begin{enumerate}
\item Using the same fitting formula Eq.~(\ref{eq:fit_1}),
but excluding the contribution from $x=0$ for all ensembles.
\item Using a different fitting formula which includes an additional $\mathcal O (a^4)$ term:
\begin{eqnarray}
m_\pi \Delta m_{\pi}(a^2, m_\pi)&=&m_{\pi,\text{phys}} \Delta m_{\pi}(0, m_{\pi,\text{phys}}) \\ \nonumber
&& \hspace{-1cm} \times\Big( 1+ c_{1}a^{2} + c_{1}|c_1|a^{4}  + c_{2}({m_{\pi}}^2-m_{\pi,\text{phys}}^{2})\Big),
\end{eqnarray} 
where the magnitude of the $\mathcal{O}(a^{4})$ term is assumed to equal to the square of the $\mathcal{O}(a^{2})$ term as an estimate of the remaining systematic effects. 
\end{enumerate}

We obtain the differences between the results of the above fits and the original fits.
The maximum of the two differences is used as the estimation of the
remaining discretization systematic error.

After continuum extrapolation,
we use the differences between 32D and 24D to correct the finite volume effects.
The absolute size of the correction is used as the estimation of
the remaining finite volume systematic error.
The continuum extrapolated, finite volume corrected results are shown in Table~\ref{tab:diagramesults} and Table~\ref{tab:diagramesults_dsdr}.
The discretization and finite volume systematic errors are combined in quadrature.
As expected, the continuum extrapolations
from the I-DSDR ensembles (Table~\ref{tab:diagramesults_dsdr})
have larger discretization systematic errors due to larger lattice spacings.
Therefore, we use the continuum extrapolation
from the Iwasaki ensembles (Table~\ref{tab:diagramesults})
as our main results.

\begin{table}[]
\begin{tabular}{|c|l|l|l|}
\hline
 & Disc (MeV) & Conn (MeV)  & Total (MeV)\\ \hline
Feyn & $0.051(9)(22) $ & $4.483(40)(28) $ & $ 4.534(42)(43) $ \\ \hline
Coul & $0.052(2)(13)$ & $4.508(46)(42) $ & $4.560(46)(41) $ \\ \hline
Coul-t & $0.018(1)(4)$ & $1.840(22)(39) $ & $1.858(22)(41) $ \\ \hline
\end{tabular}
\caption{
The continuum results from the two Iwasaki ensembles 48I and 64I.
Finite volume corrections calculated with the difference of the 32D and 24D ensembles
are already included.
The second column and third column show the quark disconnected and connected diagrams'
contributions respectively.
The bottom row shows the Coulomb potential contribution.
The statistical and systematic errors are shown respectively in the first and second set of parentheses.
}
\label{tab:diagramesults}
\end{table}

\begin{table}[]
\begin{tabular}{|c|l|l|l|}
\hline
 & Disc (MeV) & Conn (MeV)  & Total (MeV)\\ \hline
Feyn & $0.035(12)(21) $ & $4.671(49)(99) $ & $ 4.706(50)(106) $ \\ \hline
Coul & $0.050(3)(13)$ & $4.703(57)(158) $ & $4.753(58)(160) $ \\ \hline
Coul-t & $0.016(2)(4)$ & $1.931(32)(157) $ & $1.947(32)(160) $ \\ \hline
\end{tabular}
\caption{
Similar to Table~\ref{tab:diagramesults}
but extrapolated to the continuum limit with the coarser I-DSDR ensembles 24D, 32Dfine, and 24DH.
}
\label{tab:diagramesults_dsdr}
\end{table}

\begin{figure}
    \centering
    \includegraphics[scale=1]{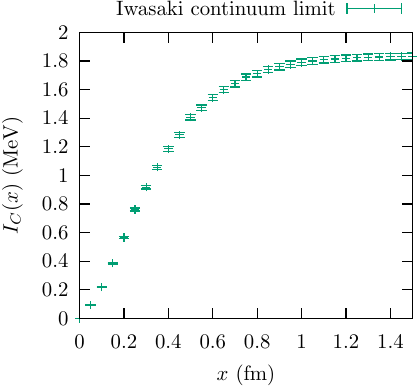}
    \caption{
    \label{fig:cgi}
    The Coulomb potential contribution to the pion mass difference.
    The curve is the partial sum, $I_{C}(x)=\frac{1}{2}\int\!\! dt\int_{|\vec{y}|\le x}\!\! d^3\vec{y}\, \mathcal{H}^L_{t,t}(t,\vec y) S^{\gamma}_{t,t}(t, \vec y)$.
    Error bars are statistical only.
    We use the results from the 48I and 64I ensembles to obtain the continuum limit,
    and include the finite volume corrections from the 32D and 24D difference.
    }
\end{figure}

As a byproduct of the calculation,
we plot the Coulomb potential contribution as a function of spatial separation at $t=0$ in Figure~\ref{fig:cgi}.
This plot provides some indication of the size and the shape of the pion.

\section{Conclusion} 

We have used the IVR method to calculate the pion mass splitting to order
$\mathcal O (\alpha_\text{QED}, (m_u - m_d) / \Lambda_\text{QCD})$ on the lattice.
The calculation is directly performed at the physical pion mass ($m_\pi = 135~\mathrm{MeV}$),
with different lattice spacings and lattice sizes.
Both the connected and the disconnected diagrams are included.
We have obtained the continuum and infinite volume limit results with Feynman gauge:
$\Delta m_\pi = 4.534(42)(43)~\mathrm{MeV}$, where the quantities in the first and second sets of parenthesis correspond respectively to
the statistical and systematic uncertainties.
We have also performed the same calculation using Coulomb gauge.
Results in both gauges are consistent with each other and
with the experimental value $4.5936(5)~\mathrm{MeV}$~\cite{Zyla:2020zbs}.
Such high precision agreement demonstrates the success of
both lattice QCD and the IVR methods.
We plan to use the IVR method in future lattice QCD + QED
spectroscopy calculations.

\begin{acknowledgements}
\section{Acknowledgements}
We would like to thank the RBC/UKQCD for the ensembles they have supplied and useful discussion many of the members therein have provided.
%
We also thank Prof. F.-K. Guo for the useful communications.
L.C.J. acknowledges support by DOE Office of Science Early Career Award DE-SC0021147 and DOE grant DE-SC0010339.
X.F. is supported in part by NSFC of China under Grants No. 11775002, No. 12125501 and No. 12070131001 and National Key Research and Development Program of China under Contracts No. 2020YFA0406400.
We developed the computational code based on
the Columbia Physics System (https://github.com/RBC-UKQCD/CPS)
and Grid (https://github.com/paboyle/Grid).
The computation is performed under the ALCC Program of the US DOE on the Blue Gene/Q (BG/Q)
Mira computer at the Argonne Leadership Class Facility, a DOE Office of Science Facility
supported under Contract DE-AC02-06CH11357.
Computations for this work were carried out in part on facilities of the USQCD Collaboration, which are funded by the Office of Science of the U.S. Department of Energy.
\end{acknowledgements}

\bibliography{sample}

\end{document}

%% file: values2.tex
\newcommand\daf{4.283(21)}
\newcommand\dac{4.375(25)}
\newcommand\cgia{1.884(13)}
\newcommand\dbf{4.415(14)}
\newcommand\dbc{4.459(15)}
\newcommand\cgib{1.875(6)}
\newcommand\dcf{3.598(12)}
\newcommand\dcc{3.825(13)}
\newcommand\cgic{1.866(7)}
\newcommand\ddf{4.002(18)}
\newcommand\ddc{4.109(21)}
\newcommand\cgid{1.863(12)}
\newcommand\dEf{3.632(10)}
\newcommand\dEc{3.823(12)}
\newcommand\cgiE{1.872(6)}
\newcommand\dff{2.406(37)}
\newcommand\dfc{2.509(12)}
\newcommand\cgif{1.498(8)}